# Neutrinoless double-beta decay of $^{76}$Ge and $^{130}$Te. A correction of the neutrinoless 2β-decay model and a reanalysis of QUORICINO results.

I.V.Kirpichnikov

SSC RF "Institute for Theoretical and Experimental Physics", Moscow

NRC "Kurchatov Institute"

An abstract

A correction of the neutrinoless 2β-decay model was proposed which predicted a shift of the 2β0ν-signal from the Q value ( 2β-decay energy ). The shifts were calculated for $^{76}$Ge ( ΔE= –2.6 keV ), $^{100}$Mo ( ΔE= –4.7 keV ), $^{130}$Te ( ΔE= –3.7 keV ). The calculations for $^{76}$Ge were in agreement with H-M results. A reanalysis of the published Quoricino data ( a search for a 2β0ν-decay of $^{130}$Te ) was performed. It indicated a presence of a signal at E = 2523.5 keV≈ Q($^{130}$Te)–3.7 keV. It could be attributed to the $^{130}$Te neutrinoless 2β-decay with $T_{1/2}$ = (5.7±2.3)·10$^{23}$ y.

A comparison of the H-M and Quoricino data has strongly supported the experimental observation of the 2β0ν-decay process and a validity of the proposed correction.

Investigations of the neutrinoless double beta-decay were intensively performed within several decades. But the only claim for the observation of the process (2β0ν-decay decay of $^{76}$Ge) was published by a group from the Heidelberg–Moscow Collaboration **[1,2,3]**. In the background spectra of Ge detectors they found a peak at (2038.07±0.44) kev, close to the energy Q=2039 keV of the 2β0ν-decay of $^{76}$Ge.

However the interpretation of the peak as a signature of the 2β0ν-decay of $^{76}$Ge seemed dubious. The analysis of the [1-3] results indicated that the 2038 kev line had a complex origin [4]. It was produced by an overlapping of three unresolved peaks: ~2035.5 kev, ~2037.5 kev and ~2039.1 kev. The 2035.5 kev and 2039.1 kev peaks were due to double-coincidences of gamma-quanta in the detectors. Only the 2037.5 kev line could be an expected signal of a 2β0ν-decay of $^{76}$Ge.

Pulse-shape analysis of the data provided possibility to pick-out the 2037.5 keV line [2,3]. Still an observed ≈ –1.5 keV shift of the peak position relative to the Q-value was a serious argument against the claim. The Q-value of the 2β0ν-decay of $^{76}$Ge was known with a very high accuracy as E=2039.0±0.005 keV. Attempts to explain the shift through law statistics or calibration uncertainties failed. It was necessary to understand a nature of this shift.

A corrected model of the 2β0ν-decay

First it has to be mentioned, that the signal *should be shifted relative the Q-value in any case*. A product nuclei emitted two electrons and nothing else. The new-born electrons took away some momentums, $p_1$ and $p_2$, ~ several MeV/c each ones. A proper recoil momentum got a product nuclei. The electrons lost a part of their energies $\Delta E_1 \sim [(p_1)^2 + (p_2)^2] / 2·M(A)$, where M(A) was a mass of a product nuclei. A loss should be about ΔE~0.02 keV and couldn't be observed.

It was valid in a case if the momentums were accepted by a product nuclei as a whole one. But was it so? Two neutrons decayed and two pairs (a proton + an electron) were born. It was quite reasonable to suppose that the new-born electrons would share *virtually* their momentums just with the new born protons. It would be possible as the 2β0ν-decay process included a virtuality.

A loss of energies by electrons would be much more essential in this version of the decay :

$$\Delta E = M(A) · \Delta E_1 = [(p_1)^2 + (p_2)^2] / 2·m(p),$$ where m(p) *was now the mass of a proton*.

The ΔE would be about several keV and could be fixed with germanium detector.

2β0ν-decay of $^{76}$Ge.

Let us consider the most probable variant of the 2β0ν-decay of $^{76}$Ge. Two electrons had equal energies t=2039 keV/2=1.0195 Mev. Their momentums were $p(e)^2 = t^2 + 2·m(e)·t = 2.081$ MeV$^2$/c$^2$. Sum loss of energies of the two electrons was ΔE=2·p(e)$^2$/2·m(p) = 2.081/938=2.22 кэв.

This value had to be corrected due to a distribution of single electron energies in the 2β0ν-decay process. The correction was rather small (~18 %) and the final shift of the signal was ΔE=–2.6 keV. The value of shift could be calculated with a rather high accuracy as it was calculated directly and not as a difference of big numbers.

The calculated shift was in agreement with the result of H.Klapdor [3]. It was also even in better agreement with the GERDA results [5] if one supposed that four events at E=2036.5 MeV were not background events (fig.1).

The above hypothesis removed a very serious argument against a claim for an observation of 2β0ν-decay of $^{76}$Ge [3]. Moreover, a coincidence of the predicted shift of the line ΔE= −2.6 keV with the observed value supported it. Doubts did exist still, first of all due to a high density of background gamma-peaks in the measured spectrum. One had to have an independent confirmation of the Klapdor claim. And such a confirmation was presented by a reanalysis of the Quoricino published data [6,7,8].

Quoricino results.

A project "Quoricino" was devoted to a search for the 2β0ν-decay of $^{130}$Te and was performed in LNGS, Italy. The first results were presented at the Neutrino-2004, Paris, by E.Fiorini [6]. The final publication was in 2011 [8]. Bolometric detectors of $TeO_2$ (source = detector) were used with parameters close to those of the Heidelberg-Moscow device. The total mass of 62 detectors was 44.7 kg, which contained 34% of active $^{130}$Te isotope (active mass ~11 kg). Energy resolution of the device was 7.2 kev FWHM at 2.6 MeV. The last publication [8] presented a total collected statistics 19.75 kg y. A signal of the 2β0ν-decay was searched at energies close to Q value. No trace of the signal was found and only a limit for the process $T_{1/2}>2.8·10^{24}$ was claimed (90% CL).

The proposed corrections stimulated a reanalysis of the published Quoricino data. A shift of the 2β0ν-decay signal of $^{130}$Te was calculated just as for $^{76}$Ge. A predicted value was ΔE= −3.7 кэв. There was a peak in Quoricino spectra [7,8] just at the energy E=2023.5 keV ≈ (Q-3.7) keV with a statistical deviations about 2σ (fig.2). An existing of a peak just at the predicted energy has strongly supported the proposed corrections. A more definite conclusion from these data seemed difficult to get because any model was absent for a Quoricino background in the ROI. The only was known about it that ~40% of a background was due to the 2.6 MeV peak [8]. All the other had to be attributed to a "flat" component from an alpha background.

Fortunately an additional information could be extracted from the early E.Fiorini report at Neutrino-2004[6]. By some reasons these data were not included in the publications of 2007 y [7] and 2011 y [8] and could be treated separately. Collected statistics was only 5.8 kg y. But an estimated background in ROI was ~ 5.0 times less that in the all further exposures due to the less intensities of the both main components. Again there was a line at the 2023.5 keV with the same intensity as in other sets (fig.3,4). The spectrum indicated also a possible presence of several background peaks.

To get a value of a life-time it was necessary to estimate a number of events in the 2023.5 keV peak. As wings of the peak drawn within a background (no model !), an intensity of the line was estimated through a single 2023 keV interval (table 1):.

| Data | 2011+2004 | 2011 | 2008 | 2004 |
|---|---|---|---|---|
| Mt[kg y] | 25.55 | 19.75 | 11.83 | 5.8 |
| ∑(2525-74)keV | 566 | 533 | 341 | 33 |
| ∑/50 = <bacgd> | 11.32 | 10.66 | 6.82 | 0.66 |
| N(2523) =1 keV | 23±4.8 | 20±4.5 | 15±3.9 | 3 |
| **Effect= 1 keV** | **11.68±4.8** | **9.34±4.5** | **8.18±3.9** | **2.34** |
| Eff / Mt | 0.457 | 0.472 | 0.691 | 0.403 |
| $T_{1/2}$ | 1.27x10$^{24}$ | 1.23x10$^{24}$ | 0.84x10$^{24}$ | 1.37x10$^{24}$ |

The corresponding life-times were given in the last line of the table. These values were evidently overestimated and could be accepted only as an upper limit for a life-time.

The width of the line seemed noticeably less then the widths of the normal gamma-peaks [fig.2-5]. It could be connected with an origin of the 2β0ν-signal. Two electrons directly generated the signal at the same point of a crystal. There were no intermediate agents – several gamma-quanta – and a scattering of points of interactions.

An attempt was made to get a more realistic shape of the line. The data of 2007 [7] were used for this purpose. The background spectrum around the line was reproduced as a sum of four components : the calculated component due to 2.6 MeV peak [8], sum $^{60}$Co line and two additional peaks at 2017 keV and 2027 keV (all with normal widths, fig.5). Parameters of the peaks were chosen rather arbitrary to reproduce the experimental piece of the spectrum. The normal resolution of the Quoricino was 7.2 keV. A width of the 2523 keV line was found some about 2 keV. A square of the line was 18.5 events compare to 8.2 events at 2523 keV single interval. One could suppose the same ratio R=18.5/8.2=2.3 for the final result 2004[6]+2011[8] .

The proper life-time would be $T_{1/2} = (5.7\pm2.3)\cdot10^{23}$ years. Still one should take in the mind that this value could be overestimated as it was calculated with the minimum width of the peak .

There were a claim for a search for 2β0ν-decay of $^{100}$Mo with bolometric detector [9]. A calculated shift of the signal for $^{100}$Mo was ΔE=4.7 keV. The measurements would provide an additional and independent check of the proposed hypothesis.

References.

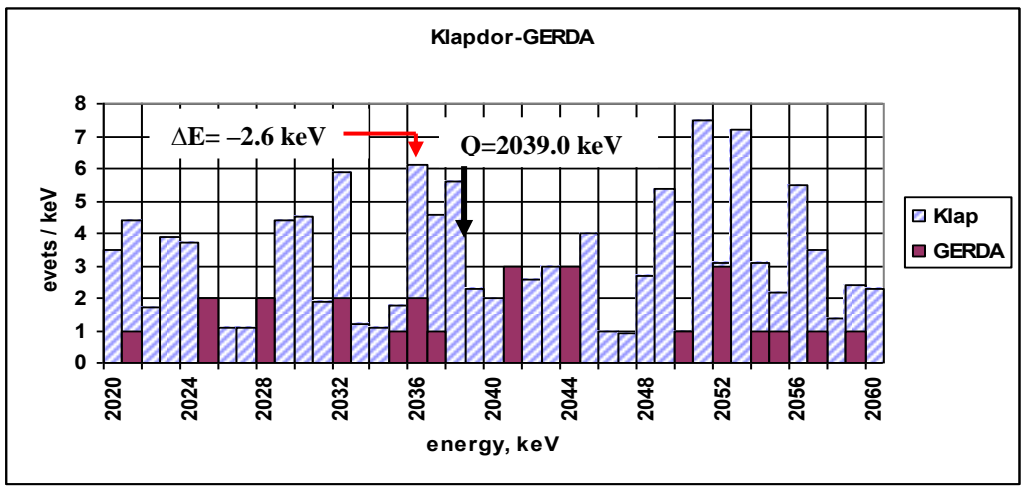

Fig.1. A comparison of Klapdor-HM and GERDA results.
K-HM results : the SSE spectrum (Mt=51.39 kg y). A shift of the peak relative to the Q-value was ΔE=–1.5 keV . A calculated shift was within an uncertainty of a calibration. A position of the peak was E = [2037.5±1.0(stat)±0.5(syst)] keV [2,3].
GERDA : full data (Mt=21.6 kg y). Four events just at the predicted energy were attributed to a background.

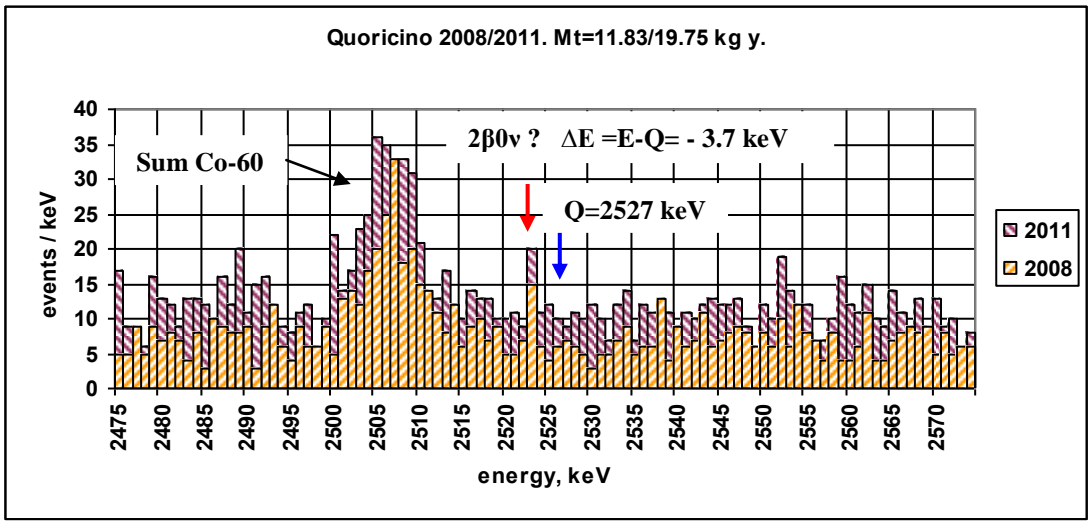

Fig.2. The experimental data of Quoricino. 2008 y (11.83 kg y [7]) and 2011 y (19.75 kg y [8]). An energy resolution of the detector at the sum $^{60}$Co peak was some worse in 2011 year.

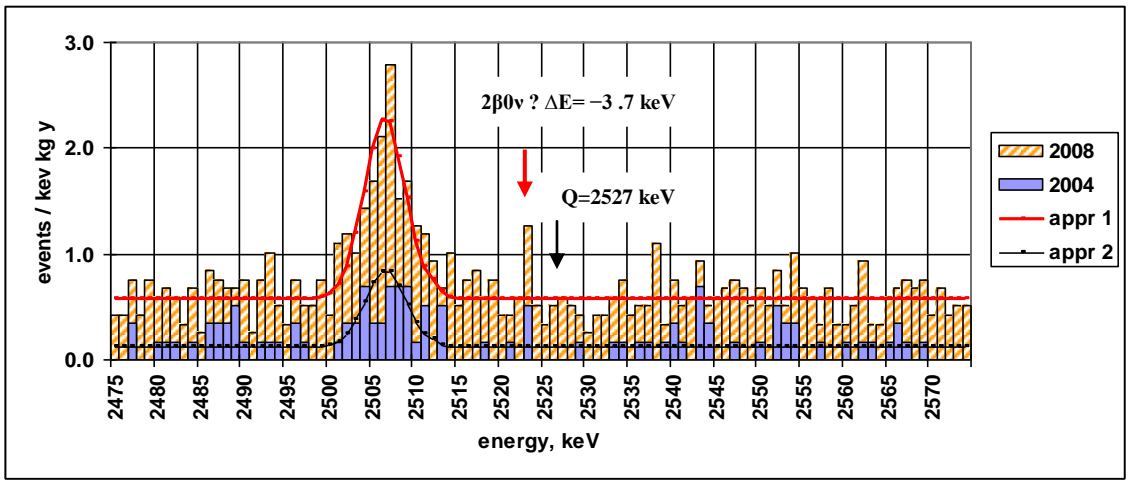

Fig.3. A comparison of the 2004 year [6] and 2008 year [7] Quoricino data.
The data were normalized through Mt values (5.8 kg y [6] and Mt=11.83 kg y [7]).
A significant change of the background was indicated. An intensity of the 2.6 MeV peak in the 2004 y data was ~5.5 times less and an intensity of the "flat" component ~4.9 times less then in all further exposures. Flat components were calculated as a mean background (peak at E=2506 keV was excluded). The approximation 2 was given by E.Fiorini.

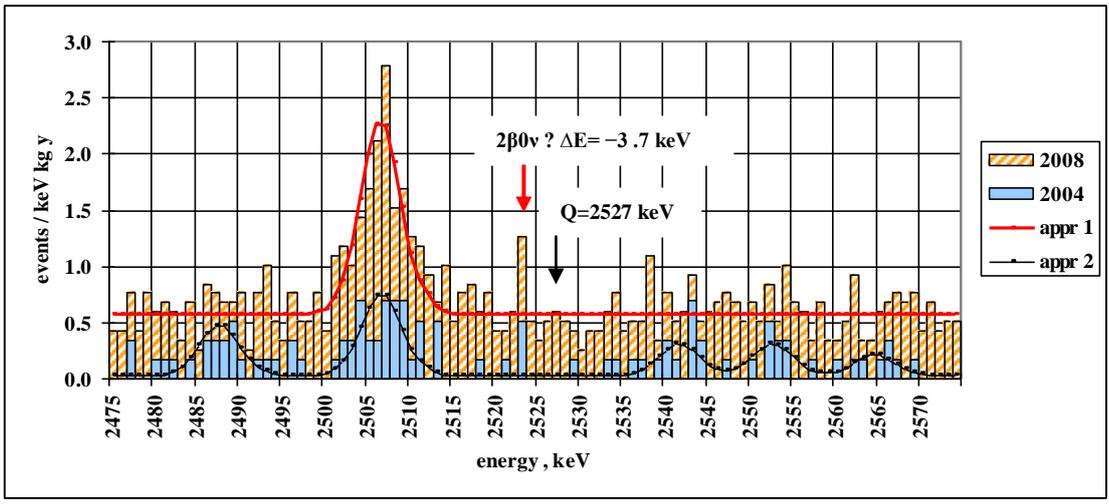

Fig.4. The same data as at fig.3. Approximation 1: a flat component was calculated as a mean background. Approximation 2 took into account possible background levels. A flat component was calculated as belonging to the 2.6 MeV peak (according [8] with a correction for an intensity of the peak).

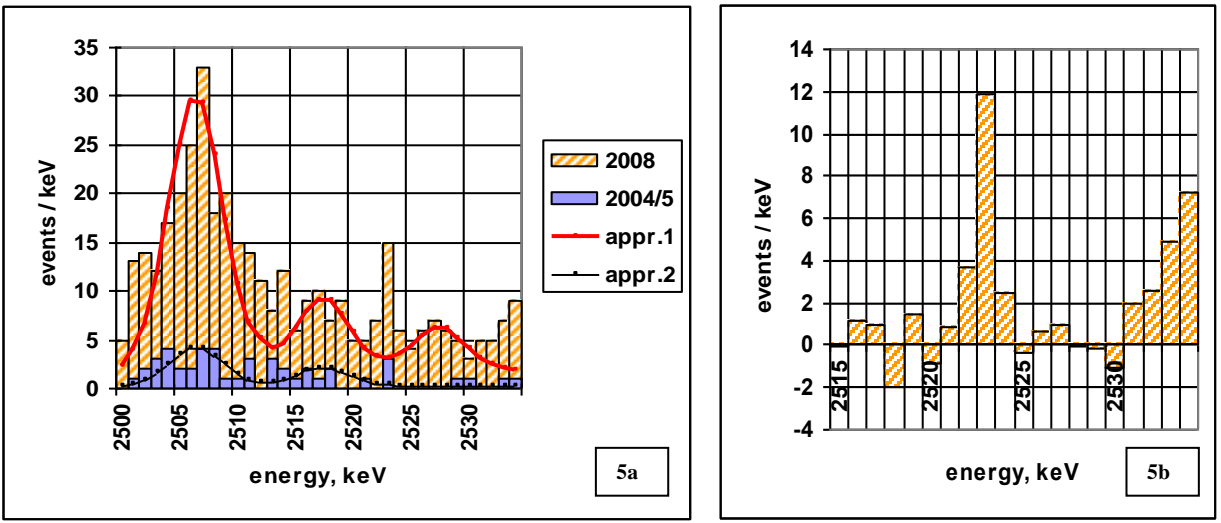

Fig.5. An attempt to get a shape of the 2023 keV line.
5a. A lower part : the data 2004/5 [6b] were used instead of [6a]. It included some other choice of exposures. The main difference was an existence of a bunch of events (11 counts) at 2017 keV - an extra gamma line ? A presence of this gamma-line was indicated also in the 2008 y data.
The approximations 1 took into account a possible background level at 2027 keV . Parameters of the peaks were chosen rather arbitrary to reproduce the experimental piece of the spectrum. Flat components were calculated according [8] as belonging to the 2.6 MeV peak.
5b. The difference "data 2008 minus approximation 1"
S(peak) = 18.5 events/5 ch(##2521-2525)